\documentclass[12pt,draft]{amsart}
\usepackage{psfig,epsf,amssymb,amsmath,latexsym}
\title[Semicircle law for random matrices with correlations]
{Semicircle law and freeness for random matrices with symmetries or
correlations }
\author{Jeffrey H.\ Schenker}
\author{Hermann Schulz-Baldes}

\address{Institut f\"ur Theoretische Physik, ETH, Z\"urich, Switzerland}
\address{Mathematisches Institut, Universit\"at
Erlangen, Germany}

\date{January 31, 2005}

\newtheorem{theo}{Theorem}

\newtheorem{lemma}{Lemma}

\newcommand{\CC}{{\mathbb C}}
\newcommand{\NN}{{\mathbb N}}
\newcommand{\RR}{{\mathbb R}}

\newcommand{\Aa}{{\mathcal A}}

\newcommand{\Pp}{{\mathcal P}}

\newcommand{\EE}{{\bf E}}

\newcommand{\Tr}{\mbox{\rm Tr}}

\newcommand{\Nn}{{\mathcal{N}}}

\begin{document}

%%%%%%%%%%%%%%%%%%%%%%%%%%%%%%%%%%%%%%%%%%%%%%%%%%%%%%
\begin{abstract}
For a class of random matrix ensembles with correlated matrix elements, it is
shown that the density of states is given by the Wigner semi-circle law. This
is applied to effective Hamiltonians related to the Anderson model in
dimensions greater than or equal to two.
\end{abstract}
%%%%%%%%%%%%%%%%%%%%%%%%%%%%%%%%%%%%%%%%%%%%%%%%%%%%%%

\maketitle

%%%%%%%%%%%%%%%%%%%%%%%%%%%%%%%%%%%%%%%%%%%%%%%%%%%%%%
\section{The Result}
It is a classical theorem due to Wigner \cite{Wig} that the density of states
of a growing sequence of real symmetric matrices with independent entries
converges in distribution to the semi-circle law. More precisely, this means
the following: Consider an $n\times n$ random matrix
$X_n=(\frac{1}{\sqrt{n}}\,a_n(p,q))_{1\leq p,q\leq n}$ where, apart from the
symmetry condition, the entries $a_n(p,q)$ are independent centered random
variables with unit variance (and a growth condition on their moments). Then
the expectation value of the moment of $X_n$ of order $k\geq 0$ satisfies
\begin{equation}
\label{eq-convergence}
\lim_{n\to\infty} \EE\;\frac{1}{n}\;\Tr_n(X_n^k)
\;=\;
\left\{
\begin{array}{cc}
C_{\frac{k}{2}}\;=\;
\frac{k !}{\frac{k}{2}! (\frac{k}{2}+1)!} &
\;\;\;\;\; k \mbox{ even }, \\
& \\
0 & \;\;\;\;\; k \mbox{ odd }.
\end{array}
\right.
\end{equation}
The coefficients $C_{k}$ are called the Catalan numbers, and the terms on the
r.h.s.\ of (\ref{eq-convergence}) are precisely the moments of the semi-circle
law, namely the absolutely continuous measure supported by the interval
$[-2,2]$ with density $\frac{1}{2\pi}\sqrt{4-x^2}$.

In this note we address the question: How many independent matrix elements are
needed such that the above asymptotic behavior still holds? In other words, how
many correlations may the matrix elements have? This is similar to an extension
of the classical central limit theorem to sums of random variables with
decaying correlations. We consider here a particular class of such
correlations, including cases with supplementary symmetries of the matrix
elements (apart from $a_n(p,q)=a_n(q,p)$). This is motivated by applications,
in particular a simplified effective Anderson model. For the two-dimensional
effective Hamiltonian, the so-called ``flip-matrix'' model introduced in
\cite{P}, it was shown in \cite{BMR} that the density of states is
semi-circular using supersymmetric functional integrals. Bellissard has
conjectured \cite{B} that this result also holds for the effective Hamiltonians
in higher dimensions.  As a consequence of our result we obtain a proof of this
conjecture (see Section~\ref{examples} for details). In Section~\ref{freeness},
we indicate that known results on asymptotic freeness \cite{Voi,Spe} extend to
the present ensembles. This  allows to calculate higher correlation functions.

The precise set-up we consider is as follows. For each $n\in\NN$, suppose given
an equivalence relation $\sim_n$ on pairs $P=(p,q)$ of indices in
$\{1,\ldots,n\}^{\times 2}$. The entries of the matrix
$X_n=(\frac{1}{\sqrt{n}}\,a_n(p,q))_{1\leq p,q\leq n}$ are complex random
variables, with $a_n(p_1,q_1),\ldots,a_n(p_j, q_j)$ independent whenever
$(p_1, q_1),\ldots,(p_j, q_j)$ belong to $j$ distinct equivalence classes of the
relation $\sim_n$. Just as above, the entries are supposed to be centered, to
have unit variance, and to obey a moment condition, namely
\begin{equation}
m_k \; := \: \sup_n\;\max_{p,q=1,\ldots,n} \;\EE(|a_n(p,q)|^k)  \; < \; \infty
\; , \label{eq-moments}
\end{equation}
for all $k\in\NN$. For equivalent pairs $(p,q)\sim_n(p',q')$, the relation
between $a_n(p,q)$ and $a_n(p',q')$ is not specified, and these variables may
be correlated. For instance, we might have $a_n(p,q) = a_n(p',q')$, describing
an auxiliary symmetry as in \cite{P,BMR,B}, but this is not necessary. However,
the following conditions are imposed on the equivalence relation:
\begin{align}
\tag{C1}  &\max_{p} \#\{(q,p',q')\in\{1,\ldots,n\}^{\times
3}\,|\,(p,q)\sim_n(p',q')\} \,=\,o(n^2)  \\ \tag{C2}
  &\max_{p,q,p'}
\#\{q'\in\{1,\ldots,n\}\,|\,(p,q)\sim_n(p',q')\} \,\leq\, B\; \intertext{for
some constant $B < \infty$, and}
\tag{C3} & \# \{ (p,q,p')
\in\{1,\ldots,n\}^{\times 3} \, |\, (p,q)\sim_n(q,p') \, \&
\, p \neq p' \} \,
= \, o(n^2)\;.
\end{align}
Finally, we require that
\begin{equation}
  a_n(p,q) = \overline{a_n(q,p)} \; , \label{eq-hermitian}
\end{equation}
so that $X_n$ is Hermitian. Thus, $(p,q) \sim_n (q,p)$ for every $p,q$, which
is consistent with (C1) to (C3).

%%%%%%%%%%%%%%%%%%%%%%%%%%%%%%%%%%%
\begin{theo}
\label{theo1} The density of states of an ensemble of Hermitian random matrices
$X_n$ obeying the moment bound \eqref{eq-moments} and associated to equivalence
relations $\sim_n$ satisfying {\rm (C1), (C2)} and {\rm (C3)}
is a semicircle law, that is, {\rm (\ref{eq-convergence})} holds.
\end{theo}
%%%%%%%%%%%%%%%%%%%%%%%%%%%%%%%%%%%

\noindent \emph{Remark:} The first condition (C1) means that the size of the
equivalence classes, \textit{i.e.}, the number of entries correlated with a
given entry of the matrix, cannot grow too fast. The second condition (C2) is
more technical, but plays a key role in the proof. It could be relaxed,
replacing $B$ by $O(n^\varepsilon)$ for every $\varepsilon > 0$, provided (C1)
were strengthened, replacing $o(n^2)$ by $o(n^{2-\varepsilon})$ for all
$\varepsilon > 0$. The third condition (C3) ensures, among other things,
that there are not too
many dependent rows, which could lead to many eigenvectors with eigenvalue
zero. (An example illustrating this phenomenon is described below.)

Our proof of this theorem is an extension of Wigner's original proof for the
fully independent case \cite{Wig,Spe,HP}, the difference being that we use an
equivalence relation on the pairs of indices rather than on the indices
themselves (which may make Wigner's proof even more transparent to some
readers).

\subsection*{Proof of Theorem \ref{theo1}}To begin, let us call a
sequence $(P_1,\ldots,P_k)$ of pairs
``consistent'' if neighboring pairs $P_l=(p_l,q_l)$ and
$P_{l+1}=(p_{l+1},q_{l+1})$ satisfy $q_l=p_{l+1}$, where $l=1,\ldots, k$ and
$k+1$ is cyclically identified with $1$. Then, for any given $k\in\NN$, we have
\begin{equation}
\label{eq-writeout}
\EE\;\frac{1}{n}\,\Tr_n(X_n^k)
\;=\;
\frac{1}{n^{1+\frac{k}{2}}}
\;
\sum_{P_1,\ldots,P_k=(1,1)}^{(n,n)}
\EE(a_n(P_1)\dots a_n(P_k))
\mbox{ , }
\end{equation}

\noindent where the sum runs over consistent sequences only. The terms of this
sum are uniformly bounded as $n \to \infty$:
$$
\left | \EE(a_n(P_1)\dots a_n(P_k)) \right | \; \le \; m_k \;,
$$
as follows from (\ref{eq-moments}) and the H\"older inequality.

To group similar terms, let us associate a partition $\pi$ of $\{1,\ldots,k\}$
to each sequence $(P_1,\ldots,P_k)$ by means of
$$
l\sim_\pi m \qquad \Longleftrightarrow \qquad P_l\sim_n P_m \;,
$$
and refer to $(P_1, \ldots, P_k)$ as a ``$\pi$ consistent sequence''.
Introducing the notation $S_{n}(\pi)$ for the set of $\pi$ consistent sequences
with indices in $\{1,\ldots,n\}$, equation (\ref{eq-writeout}) becomes
\begin{multline}
\label{eq-writeout2} \EE\; \frac{1}{n}\,\Tr_n(X_n^k) \\ \;=\;
\frac{1}{n^{1+\frac{k}{2}}} \; \sum_{\pi\in\Pp(k)}\; \sum_{(P_1,\ldots, P_k) \in
S_{n}(\pi)}\; \EE(a_n(P_1) \cdots a_n(P_k)) \;,
\end{multline}
where $\Pp(k)$ denotes the set of all partitions of $\{1, \ldots, k\}$.

Let $\#\pi$ be the number of blocks of the partition $\pi$. If
$\#\pi>\frac{k}{2}$, then there has to be one singleton block, namely a block
with only one element. Thus for any $\pi$ consistent sequence the corresponding
entry in $a_n(P_1)\cdots a_n(P_k)$ appears only once. As this variable is
centered and independent of all others appearing,  $\EE(a_n(P_1) \cdots
a_n(P_k))=0$. Therefore the sum in (\ref{eq-writeout2}) can be restricted to
partitions with less than or equal to $\frac{k}{2}$ blocks.

Next we argue that partitions with $r=\#\pi<\frac{k}{2}$ give vanishing
contribution in the limit $n\to\infty$. Since the corresponding term in the sum
over $\Pp(k)$ in (\ref{eq-writeout2}) is bounded by $m_k \times \#S_{n}(\pi)$,
it suffices to derive an upper bound on the number of $\pi$ consistent
sequences. Let us begin by choosing the pair $P_1$. There are $n^2$ choices to
be made, corresponding to $n$ choices for each of the indices. Now, the first
index of $P_2$ is fixed due to consistency. If $2 \sim_\pi 1$, then the second
index can take at most $B$ values due to condition (C2), otherwise it is
unconstrained, and can take at most $n$ values. Similarly, once we get to
$P_l$, $l<k$, we have either at most $B$ or at most $n$ possible choices for
the second index, depending on whether $l\sim_\pi j$ for one of
$j=1,\ldots,l-1$. For the last pair $P_k$ there is no freedom, due to
consistency. This shows
$$
\#S_{n}(\pi) \;\leq\; n^2\,n^{r-1}\,B^{k-r}\;.
$$
Since $r<\frac{k}{2}$, this contribution is negligible compared to the
prefactor $1/n^{1+k/2}$ in (\ref{eq-writeout2}) in the limit $n\to\infty$. In
particular, this implies that the limit in (\ref{eq-convergence}) vanishes if
$k$ is odd.

For even $k$, we may now focus on the contributions coming from partitions with
exactly $k/2$ blocks. Moreover, if the partition has a singleton the
contribution vanishes as described above. Hence the partition has to be a pair
partition, \textit{i.e.}, each block has exactly two elements. (Note that the
``pair'' in ``pair partition'' has nothing to do with the fact that each point
$l$ is
associated to a pair $P_l$ of indices.)%

Let $\Pp\Pp(k)$ denote the pair partitions of $k$. To prove Theorem
\ref{theo1}, we must control somewhat more carefully the asymptotics in $n$ of
$\#S_{n}(\pi)$ for a pair partition $\pi \in \Pp\Pp(k)$. This will be
accomplished in the following four lemmas.

%%%%%%%%%%%%%%%%%%%%%%%%%%%%
\begin{lemma}
\label{lem1} Let $\pi\in\Pp\Pp(k)$ be a pair partition containing a pair of
neighbors, that is $m\sim_\pi m+1$ for some $m$, and let $\pi'\in\Pp\Pp(k-2)$
be the partition obtained by eliminating the corresponding pair {\rm
(}and relabeling
$l \mapsto l -2$ for $m+2 \le l \le k${\rm )}. Then
\begin{equation}
\#S_{n}(\pi) \;\leq\; n \times \#S_{n}(\pi') \; + \; o(n^{\frac{k}{2} +1}) \;.
\label{eq-elimination}
\end{equation}
\end{lemma}
%%%%%%%%%%%%%%%%%%%%%%%%%%%

\begin{proof} Let us look at the situation close to $m$ and $m+1$. The
indices are $(p_{m-1},q_{m-1})$, $(p_{m},q_{m})$, $(p_{m+1},q_{m+1})$ and
$(p_{m+2},q_{m+2})$. By consistency, we have $q_m=p_{m+1}$. Now consider
separately the two cases (i) $p_m = q_{m+1}$ and (ii) $p_m \neq q_{m+1}$.

In case (i), after two applications of the consistency condition, we get
$q_{m-1}=p_{m+2}$. Hence after eliminating the pairs $P_m$ and $P_{m+1}$, we
have a $\pi'$ consistent sequence. Therefore, in this case there are $n$
choices for $q_m = p_{m+1}$ and at most $\#S_{n}(\pi')$ choices for the
remaining indices, giving the first term in \eqref{eq-elimination}.

In case (ii), by (C3) there are only $o(n^2)$ choices for the triple $p_m$,
$q_m = p_{m+1}$, $q_{m+1}$.  Since there are $(k-2)/2$ pairs remaining, there
are no more than $n^{(k-2)/2} B^{(k-2)/2 - 1}$ choices for the remaining
indices. To see this, start at $m+2$ and proceed through the remaining indices
cyclically as in the argument to eliminate partitions with $\# \pi < k/2$
above. Combining the two factors $o(n^2) \times O(n^{(k-2)/2})$ gives the
second term in \eqref{eq-elimination}.
\end{proof}

A partition $\pi$ is called crossing if there are positions
$m_1<m_2<m_3<m_4$ such that $m_1\sim_\pi m_3$ and $m_2\sim_\pi m_4$,
otherwise it is called non-crossing.

%%%%%%%%%%%%%%%%%%%%%%%%%%%%%%%%%%%%%%%%
\begin{lemma}
\label{lem2} Let $\pi\in\Pp\Pp(k)$ be a crossing pair partition, then
$$
\lim_{n\to\infty} \,\frac{\#S_{n}(\pi)}{n^{\frac{k}{2}+1}} \;=\; 0 \; .
$$
\end{lemma}
%%%%%%%%%%%%%%%%%%%%%%%%%%%%%%%%%%%%%%%%

\begin{proof} The first step is to apply Lemma \ref{lem1} as many times
as possible, eliminating all nearest neighbor pairs in $\pi$ and the resulting
reduced partitions. The end result is that
$$
\# S_n(\pi) \;\leq\; n^r\,\# S_n (\pi') \; + \;  o(n^{\frac{k}{2} +1})\;,
$$
where $r$ is the number of pairs eliminated, and $\pi'\in\Pp\Pp(k-2r)$ is
maximally crossing in the sense that it cannot be further reduced by the
procedure of Lemma \ref{lem1}, \textit{i.e.}, for every $m$ we have $m \not
\sim_{\pi'} m+1$. Because $\pi$ is crossed, there are at least two pairs left,
so $k -2 r \ge 4$. Consider a pair $m \sim_{\pi'} m+\ell$ of $\pi'$ with
minimal $\ell$, understood in a cyclic sense on the $k-2r$ positions of $\pi'$.
Then $\ell \ge 2$ and all points $j$ with $m<j<m+\ell$ form pairs with points
outside $\{m,\ldots,m+\ell\}$, \textit{i.e.}, these pairs cross the pair $m
\sim_{\pi'} m+\ell$.

To estimate $\# S_n(\pi')$, let us first count the choices for indices in the
interval $\{m,\ldots, m+\ell \}$, first choosing $P_m=(p_m,q_m)$,
$P_{m+\ell}=(p_{m+\ell}, q_{m+\ell})$, and then successively $P_{m+1}=(p_{m+1},
q_{m+1})$, $P_{m+2}, \ldots$. There are $n$ choices for $p_m$ and then, by
(C1), $o(n^2)$ choices for the three indices $q_m$, $p_{m+\ell}$, $q_{m+\ell}$.
Now, $p_{m+1}$ is fixed by consistency, but $q_{m+1}$ is unconstrained aside
from the requirement $P_{m+1} \not \sim_n P_m$, which we ignore to obtain an
upper bound. The same holds for $P_{m+2},\ldots,P_{m+\ell-2}$, however, for
$P_{m+\ell-1}$ there is no freedom due to consistency. For the whole interval
$\{m,\ldots,m+\ell\}$ this gives an upper bound $n\times o(n^2)\times
n^{\ell-2}=o(n^{\ell +1})$ on the number of choices.

Finally, let us go through the remaining $k-2r-\ell+1$ pairs of indices in
increasing order (cyclically), starting with $P_{m+\ell+1}$. Consistency fixes
the first index. If $m+\ell+1$ is paired under $\pi'$ with one of the
previously considered points, then (C2) fixes also the second index within a
set of size no larger than $B$. If $m+\ell+1$ does not form a pair with any
previous point, then there are at most $n$ choices for the second index. As we
go through the remaining $k-2r-\ell-1$ pairs, the latter case occurs exactly
$\frac{k-2r}{2}-\ell$ times. We deduce that
$$
\#S_n(\pi') \;\leq\; n^r\times o(n^{\ell +1})\times
n^{\frac{k}{2}-r-\ell}\,B^{\frac{k}{2} - r - 1}\, +\, o(n^{\frac{k}{2} + 1})
\;=\; o(n^{\frac{k}{2} + 1}) \;.
$$
\end{proof}

Thus we may restrict the sum in (\ref{eq-writeout2}) to non-crossing pair
partitions of $\{1,\ldots, k\}$, denoted  by $\Nn \Pp \Pp(k)$.  For $\pi \in
\Nn \Pp \Pp(k)$, let $PS_n(\pi)$ denote the subset of $S_n(\pi)$ made up of
$\pi$ consistent sequences such that for any two sites  $m,m'$ paired by $\pi$,
$m \sim_\pi m'$, we have $P_{m'}=(q_m, p_m)$, where $P_m=(p_m,q_m)$. Note that
for $(P_1, \ldots, P_k) \in PS_n(\pi)$ we have
\begin{equation}\label{eq-expectPSn}
  \EE (a_n(P_1) \cdots a_n(P_k)) =1 \; ,
\end{equation}
since $a_n(P_m) = \overline{a_n(P_{m'})}$ for equivalent sites $m \sim_{\pi}
m'$ and the $a_n(P)$ have unit variance.  Furthermore, there are not too many
sequences missed by $PS_n(\pi)$:

%%%%%%%%%%%%%%%%%%%%%%%%%%%%%%%%%%%%%%%%
\begin{lemma}
  \label{lem3} Let $\pi \in \Nn\Pp\Pp(k)$ be non-crossing.  Then
  $$ \lim_{n \rightarrow \infty} \frac{\# S_n(\pi) - \# PS_n(\pi)
  }{n^{\frac{k}{2}+1}} \; = \; 0\;.
$$
\end{lemma}
%%%%%%%%%%%%%%%%%%%%%%%%%%%%%%%%%%%%%%%%

\begin{proof}
Let $NS_n(\pi) = S_n(\pi) \setminus PS_n(\pi)$.  Repeating the proof of Lemma
\ref{lem2}, we see that
$$
\#NS_n(\pi) \; \le \; n \times \# NS_n(\pi')\; +\; o(n^{\frac{k}{2} +1}) \; ,
$$
where $\pi'$ is a partition of $k-2$ obtained by removing a nearest neighbor
pair from $\pi$. Referring to the notation of the proof of Lemma \ref{lem1},
the only new feature here is that in case (i), after eliminating the pair $P_m
\sim_n P_{m+1}$ (and relabeling $m+2 \mapsto m$, etc.), we get a sequence in
$NS_n(\pi')$.  This follows because the eliminated pair was of the form $(p,q)
\sim_n (q,p)$, so the ``defect'' is still present in the reduced sequence.

Iteration of the above gives
$$
NS_n(\pi) \; \le \; n^{\frac{k}{2}-1} \#  NS_n(\pi'') \;+\; o(n^{\frac{k}{2} +1})
\; ,
$$
where $\pi'' \in \Pp\Pp(2)$ is a pair partition for which $\# NS_n(\pi'') =
o(n^2)$, by condition (C3).
\end{proof}

After restricting to pair partitions in \eqref{eq-writeout2} and applying
Lemmas \ref{lem1} to \ref{lem3} and equation \eqref{eq-expectPSn}, we conclude
that
\begin{equation}
\label{eq-writeout3} \lim_{n \rightarrow \infty} \EE\;
\frac{1}{n}\,\Tr_n(X_n^k) \\
\;=\; \lim_{n \rightarrow \infty}\; \frac{1}{n^{1+\frac{k}{2}}} \;
\sum_{\pi\in\Nn \Pp\Pp(k)}\; \# PS_{n}(\pi) \; ,
\end{equation}
where, by the following lemma, the limit on the r.h.s. is in fact just
the number of non-crossing pair partitions.

%%%%%%%%%%%%%%%%%%%%%%%%%%%%%%%%%%%%%%%%
\begin{lemma}
\label{lem4} If $\pi\in\Nn\Pp\Pp(k)$, then $$ \lim_{n\to\infty}
\,\frac{\#PS_{n}(\pi)}{n^{\frac{k}{2}+1}} \;=\; 1 \; .
$$
\end{lemma}
%%%%%%%%%%%%%%%%%%%%%%%%%%%%%%%%%%%%%%%%

\begin{proof}
We first note that $\frac{k}{2} -1 $ applications of Lemma \ref{lem1} imply
$$
\#S_{n}(\pi) \;\leq\; n^{\frac{k}{2}-1}\times\#S_{n}(\pi') \; + \;
o(n^{\frac{k}{2}+1}) \;,
$$
where $\pi'\in \Pp\Pp(2)$ is a pair partition, for which $\#S_{n}(\pi')=n^2
+o(n^2)$, by (C3). Thus
\begin{equation}
  \# PS_n(\pi) \; \leq \; \#S_{n}(\pi) \; \leq \;
  n^{\frac{k}{2} +1} \; + \;  o(n^{\frac{k}{2} +1}). \label{eq-upper}
\end{equation}

In order to prove a lower bound, let us count the number of choices for
indices, proceeding from left to right. We first have exactly $n^2$ choices for
$(p_1,q_1)$.  This determines $p_2=q_1$ by consistency.  If $1 \sim_\pi 2$ then
$q_2=p_1$ is fixed since we require a $PS_n(\pi)$ sequence, otherwise there
remain exactly $n-1$ choices for $q_2$ (since $q_2 = p_1$ is forbidden because
$1 \not \sim_\pi 2$). More generally, once we reach the $j^{\mathrm{th}}$ pair
of indices, $p_j=q_{j-1}$ is fixed by consistency, but $q_j$ is fixed if and
only if $j$ is paired under $\pi$ with a number $i < j$. If $j$ is not paired
with a prior number, there remain at least $n - (j-1) B \ge n-(k-1) B$ choices
for $q_j$. Indeed this follows from (C2) since $q_j$ is free up to the
constraint that $(p_j,q_j) \not \sim_{n} (p_i, q_i)$ for $i < j$. We obtain in
this way the lower bound
\begin{equation}
\# PS_n(\pi) \;  \ge \; n^2 (n-(k-1)B)^{\frac{k}{2} -1} \; \ge \; (n-(k-1)
B)^{\frac{k}{2} +1} \; , \label{eq-lower}
\end{equation}
since there are exactly $k/2$ pairs in $\pi$, and thus $k/2+1$ free indices.

Combining the two bounds \eqref{eq-upper} and \eqref{eq-lower} completes the
proof.
\end{proof}

This concludes the proof of Theorem \ref{theo1}, because the number of
non-crossing partitions, $\# \Nn\Pp\Pp(k)$, is precisely equal to the Catalan
number $C_{\frac{k}{2}}$ \cite{Spe,HP}. \qed

%%%%%%%%%%%%%%%%%%%%%%%%%%%%%%%%%%%%%%%%%%%%%%%%%%%%%%
\section{Examples}
\label{examples}

\subsection{Flip matrix model} \label{sec-flip}
For each $n$, let $\Phi_n : \{1, \ldots,
n\} \rightarrow \{1, \ldots, n\}$ be the involution $\Phi_n(p) = n+1-p$, and
take for the equivalence relation
$$
(p,q)\; \sim_n\; (q,p)\; \sim_n\; (\Phi_n(p), \Phi_n(q))
\; \sim_n\; (\Phi_n(q),
\Phi_n(p)) \; .
$$
Clearly (C1) and (C2) are satisfied. For even $n$, (C3) holds because $(p,q)
\sim_n (q,p')$ if and only if $p=p'$. For odd $n$, this is true except for
pairs of the form $(p, \frac{n+1}{2}) \sim_n (\frac{n+1}{2}, \Phi_n(p))$, since
$\frac{n+1}{2}$ is a fixed point of $\Phi_n$. As there are only $n-1$ such
pairs, (C3) holds. Thus, Theorem \ref{theo1} applies and the density of states
is a semi-circle law.

Restricting to even $n$ and taking for the matrix elements complex Gaussians,
we obtain the flip matrix model of \cite{P}. Theorem \ref{theo1} in this
special case reproduces a result derived in \cite{BMR}. We comment
below on how this ensemble appears in connection with the Anderson model.

\subsection{Condition (C3)}
The importance of condition (C3) is illustrated by the following example. Let
the equivalence relation be induced by
$$
  (p,q) \;\sim_n\; (q,p)\; \sim_n\; (\Phi_n(p), q)
  \;\sim_n\;(\Phi_n(q),\Phi_n(p)) \; ,
$$
where $\Phi_n$ is as above. Unless $p$ or $q$ are fixed points of
$\Phi_n$, the class of $(p,q)$ contains 8 elements.
Conditions (C1) and (C2) are clearly satisfied. However, (C3) is violated since
$(p,q) \sim_n(q, \Phi_n(p))$ for every $p,q$, giving at least $n \times (n-1)$
pairs $(p,q) \sim_n(q,p')$ with $p \neq p'$. Consider, associated to this
$\sim_n$, an ensemble of real symmetric matrices with $a_n(P) = a_n(P')$ when
$P\sim_nP'$. The matrix $X_n$ has a $[ n/2]$
dimensional null space (here $[a]$ denotes the integer part of
$a\in\RR$), consisting of vectors $(v_1,\ldots ,v_n)$ with $v_j =
-v_{n+1 - j}$. As a result, the density of states for this ensemble has an atom
at $0$.  In fact, the density of states can be
computed by noting that $$X_n \; \cong \; \frac{1}{\sqrt{2}} \begin{pmatrix}   1 & 1 \\
1 & 1
\end{pmatrix} \otimes Y_{n/2} \; \cong \; \begin{pmatrix}
  \sqrt{2} & 0 \\
  0 & 0
\end{pmatrix} \otimes Y_{n/2} \; , $$
where $\cong$ denotes unitary equivalence and $Y_n$ is an ensemble of matrices
satisfying the conditions of Theorem \ref{theo1}. Thus the density of states is
a superposition of an atom at $0$ and a scaled semi-circle law:
$$
\frac{1}{2}\, \delta(x)\, \mathrm{d} x
\;+\;
\frac{1}{2}\, \frac{1}{2 \pi} \sqrt{4 - 2
x^2} \;\chi(x^2\leq 2)\,\sqrt{2}\, \mathrm{d} x \; .
$$

\subsection{Higher dimensional flip models} Let us first describe how
these models appear as effective Hamiltonians for a finite size approximation
of the Anderson model in the weak coupling limit, and then apply
Theorem~\ref{theo1} to them.

The Anderson Hamiltonian $H=H_0+ V$ on the $\ell^2$ space over a square lattice
is the sum of a translation invariant part $H_0=H_0^*$ and a (small) random
potential $V$. Consider such an operator in a finite volume
$$
\Lambda_L \; = \; \left \{-\frac{L}{2}+1,\ldots, \frac{L}{2} \right \}^d  \; ,
$$
together with periodic boundary conditions. Then $V$ is a multiplication
operator on $\ell^2(\Lambda_L)$ given by a random real-valued function $v$ on
$\Lambda_L$. Furthermore the orthonormal basis $\phi_p(x) =\mathrm{e}^{i p
\cdot x}/L^{d/2}$, with quasimomentum $p$ in the (discrete) Brillouin zone
$\Lambda_L^* = \frac{2 \pi}{L} \Lambda_L$, consists of eigenfunctions of
$H_0$ with
eigenvalues $\mathcal{E}(p)$, where $\mathcal{E}$ is some smooth periodic
function on the torus $[-\pi, \pi)^d$ --- the so-called \emph{symbol} of $H_0$.

For small $V$, the spectral analysis of $H$ near a fixed energy $E$ involves
primarily states $\phi_p$ with quasimomenta in the discrete Fermi surface
\begin{equation}\label{eq-Fermi}
\Lambda_L^*(E) \; = \; \Bigl\{ p \in \Lambda_L^*\; \Bigl|
\;|\mathcal{E}(p) - E| \le
\frac{\beta}{L} \Bigr\} \; ,
\end{equation}
with $\beta>0$ a parameter giving its width. The restriction of the random
potential to the corresponding subspace will be the higher dimensional flip
model. It is hence the lowest order in a nearly degenerate perturbation theory
for the Anderson Hamiltonian at scale $L$.

More precisely, the multiplication operator $\psi(x) \mapsto v(x) \psi(x)$ acts
on the basis $\phi_p$ as follows:
$$
  \phi_p \;\mapsto \;\sum_{q\in\Lambda^*_L}
\widehat v(p-q) \phi_{q} \; ,
$$
where $\widehat v$ is the (finite sum) Fourier transform of $v$, \textit{i.e.},
$$
  \widehat v(p) \;= \;
\frac{1}{L^d} \sum_{x \in \Lambda_L} v(x)\,
  \mathrm{e}^{i
  p\cdot x} \; ,
$$
and $p-q$ is evaluated modulo $2 \pi$. Restriction to the Fermi
surface gives thus
the following convolution operator $A_\omega$ on $\ell^2(\Lambda_L^*(E))$
$$
A_\omega \phi(p) \; = \; \sum_{q \in \Lambda_L^*(E)} \widehat v(p-q)
\phi(q) \; .
$$
Hence we are led to introduce the random matrix
\begin{equation}
\label{eq-model}
Y_n(p,q) \; = \; \widehat v(p-q) \; , \qquad p,q \in \Lambda_L^*(E) \; .
\end{equation}
Here $n= \# \Lambda_L^*(E) = O(L^{d-1})$.

If one takes $v(x)$ to be independent and identically distributed centered
Gaussians, then the variables $\widehat v(p)$ are also centered Gaussians,
which are mutually independent apart from the constraint $\overline{\widehat
v(p)} = \widehat v( -p)$ which results because the $v(x)$ are real. In terms
of the variance $\lambda^2 = \EE(v(x)^2)$, the variance of $\widehat v(p)$ is
$\lambda^2 / L^d$. The higher dimensional flip matrix model is the random
matrix $Y_n$ defined in \eqref{eq-model}, with the index set given by
\eqref{eq-Fermi}, where random variables $\widehat v(p)$ have variance
$\lambda^2 / L^d$ and satisfy the constraint $\overline{\widehat v(p)} =
\widehat v( -p)$, as well as the bounds
$$
\max_{p\in\Lambda^*_E} \;
\EE(|\widehat v(p)|^k) \; \le \; M_k\, \lambda^k\, L^{-\frac{kd}{2}} \; .
$$
To simplify the discussion, we suppose that $\mathcal{E}(p)= p^2/2$ in a
neighborhood of $E>0$, so that $\Lambda_L^*(E)$ is the intersection of a
spherical shell with the lattice $\Lambda_L^*$. This eases some arguments, but
does not affect the end result provided the level sets $\{ p\in\Lambda_L^*\,
|\, \mathcal{E}(p) =E \pm \delta\}$, for small $\delta$, are sufficiently
regular and have curvature bounded away from zero.

\noindent \emph{Remark:}
The flip matrices of Section~\ref{sec-flip}
are a simplification of $Y_n$ in $d=2$ obtained via a course
graining procedure.  The shell $\Lambda_L^*(E)$ is divided into $m$ blocks,
each subtending an angle of size $O(L^{-\frac{1}{2}})$. Each block contains
$O(L^{-\frac{1}{2}} \times L^{-1} \times L^{2}) = O(L^{\frac{1}{2}})$ points.
The flip matrix $F_m(b,b')$ is indexed by pairs of \emph{blocks}, with Gaussian
$F_m(b,b')$ which are pairwise independent except that $$F_m(b,b') \; = \;
\overline{F_m(b',b)} \; = \; F_m(-b',-b) \; = \; \overline{F_m(-b,-b')} \; .$$
The above coarse graining was introduced in \cite{P,BMR} as a simplification to
avoid the combinatorics required to deal with matrix elements $Y_n(p,q)$ with
$p-q$ small. We do not need coarse graining here, because Theorem
\ref{theo1} applies directly to an appropriate rescaling of $Y_n$.

We now turn to the analysis of the higher dimensional flip model. The matrix
elements $Y_n(p,q)$ and $Y_n(p',q')$ are statistically independent unless $ p-q
=  \pm (p'-q')$, and thus we define $\sim_n$ accordingly:
$$
(p,q) \sim_n (p',q') \;\;\; \Longleftrightarrow \;\;\;
p-q \; = \; \pm\, (p'-q')
$$
Condition (C2) is clearly satisfied, but conditions (C1) and (C3) require
proof.

Condition (C3) is a bit easier to verify. The relevant pairs $(p,q)
\sim_n(q,p')$ with $p\neq p'$ have to satisfy $p'=2q-p$. Fixing
$q\in \Lambda_L^*(E)$, this implies that the conditions
$p \in \Lambda_L^*(E)$ and $2q-p \in \Lambda_L^*(E)$ have to hold
simultaneously. The relevant set
of $p$'s is contained in the intersection of two spherical shells:
$$
\left\{
|p^2/2 -
E| \le \frac{\beta}{L} \right\}
\, \cap \,
\left\{
|(p-2q)^2/2 - E| \le \frac{\beta}{L}\right\} \; .
$$
As $|q|^2/2=E$, these shells are barely touching and there are
$$
O(L^{-\frac{d+1}{2}} L^d) \; = \; O(L^{\frac{d-1}{2}}) \; = \; o(L^{d-1})
\; = \; o(n)
$$
points in the intersection of this set with $\Lambda_L^*(E)$, essentially
because the $d$-dimensional volume of the set is $O(L^{-\frac{d+1}{2}})$ and
the lattice cell volume of $\Lambda_L^*$ is $(2\pi)^d/L^d$. As this holds for
all $n$ values of $q$, we conclude that (C3) holds.

For (C1), we need to estimate, given a fixed value of $p$,
the number of triples $q,p',q' \in
\Lambda_L^*(E)$ with $(p,q) \sim_n (p',q')$.  First note that
$$
q' \; = \; p' \, \pm \, (p-q) \; ,
$$
so that $q'$ takes 2 values once $q$ and $p'$ are chosen. Furthermore
$$
|q'|^2 - |p'|^2 \; = \;  \pm \, 2 p'\cdot (p-q) \, + \, |p-q|^2 \; .
$$
Because $p',q' \in \Lambda_L^*(E)$, this implies
$$
\left | \frac{|p-q|}{2} \pm p' \cdot \frac{(p-q)}{|p-q|} \right | \; \le \;
\frac{2 \beta}{L |p-q|} \; .
$$

In order to count the number of possible values for $q$ and $p'$,
we consider separately the cases of large and small $p-q$.
If $|p- q| \ge L^{-\frac{1}{2}}$, then
$$
\left | \frac{|p-q|}{2|p'|} \pm \cos \theta \right | \; \le \; \frac{2
\beta}{|p'| \sqrt{L}} \; ,
$$
where $\theta$ is the angle between $p'$ and $(p-q)$.  Because for all $t$
$$
|\{ \theta \in [-\pi,\pi] \, | \, |\cos \theta  - t | \le \delta \} | \; \le \;
\sqrt{2\,\delta}\;,
$$
we conclude that the allowed set of $p'$ is contained in a
spherical cone which subtends an angle no larger than $O(L^{-\frac{1}{4}})$.  The
intersection of this spherical cone with $\Lambda_L^*(E)$ contains
$O(L^{-1-\frac{1}{4}(d-1)} \times L^d) = O(L^{\frac{3}{4}(d-1)})$ points. As
there are $O(L^{d-1})$ points $q\in \Lambda_L^*(E)$
with $|p-q| \ge L^{-\frac{1}{2}}$, we end up with no
more than $O(L^{\frac{7}{4}(d-1)})$ triples in this case. However, in the other
case there are only $O(L^{-1 - \frac{1}{4} (d-1)} \times L^{d}) =
O(L^{\frac{3}{4} (d-1)})$ points $q \in \Lambda_L^*(E)$ with $|p-q| \le
L^{-\frac{1}{2}}$, and thus no more than $O(L^{\frac{3}{4} (d-1)} \times L^{d-1}) =
O(L^{\frac{7}{4}(d-1)})$ triples. Altogether we conclude that $$ \# \{
(q,p',q') \in \Lambda_L^*(E) \, | \, (p,q) \sim_n(p',q')\} \; = \;
O(L^{\frac{7}{4}(d-1)}) \; = \; o(L^{2(d-1)}) \; , $$ and thus that (C1) holds.

Therefore, Theorem \ref{theo1} applies to $X_n = c_n Y_n$ with a scalar $c_n$
chosen so that the variance of the matrix elements of $X_n$ is $1/n$, namely
$c_n^2 =(L^d/\lambda^2) \times n^{-1} =O(L)/\lambda^2$. In the context of the
Anderson model, it is natural \cite{P,B} to take $L \approx \lambda^{-2}$,
giving $n(\lambda)=\# \Lambda_L^*(E)=O(\lambda^{2-2d})$. Then
$Y_{n(\lambda)}(p,q)\approx\lambda^2 \times X_{n(\lambda)}(p,q)$ where the
$X_n$ are matrices which satisfy the hypotheses of Theorem~\ref{theo1}. It
follows that the density of states of $Y_{n(\lambda)}$ is asymptotic as
$\lambda\rightarrow 0$ to a scaled semi-circle law of width of order
$\lambda^2$.

%%%%%%%%%%%%%%%%%%%%%%%%%%%%%%%%%%%%%%%%%%%%%%%%%%%%%%%%%%%%%%%%%
\section{Asymptotic Freeness}\label{freeness}
The semicircle law is the only stable law of Voiculescu's free probability
theory for which all moments exist,
playing a role there analogous to that of the Gaussian in classical
probability. The above techniques also allow to extend known results on
asymptotic freeness of random matrices \cite{VDN,Voi,Spe,HP}. Suppose given a
family $(X^{(i)}_n)_{i\in I}$ of random $n\times n$
matrices with some index set $I$. For
different indices $i$, the entries of the matrices are supposed to be
independent, and for each $i$, the family $(X^{(i)}_n)_{n\in \NN}$ is specified
by a possibly $i$-dependent equivalence relation satisfying conditions (C1),
(C2) and (C3). Furthermore suppose given a second family $(D^{(j)}_n)_{j\in J}$
of diagonal real random matrices, the entries of which are also independent
among themselves and with respect to the family $(X^{(i)}_n)_{i\in I}$. Each
$(D^{(j)}_n)_{n\in \NN}$ is supposed to be such that a well-defined density of
states exists, namely a probability measure $\Nn_j$ such that
$$
\lim_{n\to\infty} \;\EE\,\frac{1}{n}\;\mbox{Tr}_n(f(D^{(j)}_n)) \;=\; \int
\Nn_j(dx)\;f(x) \;, \qquad f\in C_0(\RR)\;.
$$

The asymptotic behavior (in $n$) of the families $(X^{(i)}_n)_{i\in I}$ and
$(D^{(j)}_n)_{j\in J}$ now defines a state $\varphi$ on the algebra
$\Aa=\CC\bigl[\{x^{(i)}\}_{i\in I},\{d^{(j)}\}_{j\in J}\bigr]$ of
non-commutative polynomials by
\begin{multline*}
\varphi\bigl(Q(\{x^{(i)}\}_{i\in I},\{d^{(j)}\}_{j\in J})\bigr) \\ \;=\;
\lim_{n\to\infty} \;\EE\,\frac{1}{n}\; \mbox{Tr}_n\bigl(Q(\{X^{(i)}_n\}_{i\in
I},\{D^{(j)}_n\}_{j\in J})\bigr) \;, \qquad Q\in\Aa\;.
\end{multline*}

%%%%%%%%%%%%%%%%%%%%%%%%%%%%%%%%%%%
\begin{theo}
\label{theo2} The commutative subalgebras $\CC[x^{(i)}]$
and $\CC[d^{(j)}]$ are free in
$(\Aa,\varphi)$ for all $i\in I$ and $j\in J$. Moreover, the marginals of
$\varphi$ on $\CC[x^{(i)}]$
and $\CC[d^{(j)}]$ are respectively the semicircle law and
the measure $\Nn_j$.
\end{theo}
%%%%%%%%%%%%%%%%%%%%%%%%%%%%%%%%%%%

The first statement means, in particular, that all mixed moments of $\varphi$
can be calculated from the moments of the semicircle law and the densities
$\Nn_j$ by use of free combinatorics discussed in detail in \cite{Spe,HP}. In
view of the above arguments, the proof of this theorem is a straightforward
generalization of the one given in \cite{Spe}.

\noindent \emph{Acknowledgment:} We would like to thank J. Bellissard, for
bringing the flip-matrix models to our attention and explaining their link with
the Anderson model, and M. Disertori, for discussions related to this work.

%%%%%%%%%%%%%%%%%%%%%%%%%%%%%%%%%%%%%%%%%%%

\end{document}